\documentclass[a4paper,11pt]{article}

\usepackage[parfill]{parskip}
\usepackage[english]{babel}
\usepackage[utf8]{inputenc}
\usepackage{amsmath,amsfonts,fancyhdr}
\usepackage{graphicx}
\usepackage[colorinlistoftodos]{todonotes}
\usepackage[margin=1in]{geometry}
\usepackage{enumerate}
\usepackage{bigstrut}
\usepackage{breqn}
\usepackage[font=footnotesize,labelfont=bf]{caption}
\usepackage{multirow}
\usepackage{float}
\usepackage{subfig}
\usepackage{footnote}
\usepackage{hyperref}

\usepackage[final]{changes}

\usepackage{setspace}

\title{WMW Power Paper} 
\author{Name}
\begin{document}
\begin{flushleft}
\Large{\Large{\textbf{Exact Power of the Rank-Sum Test \added{for a Continuous Variable} }}}
\end{flushleft}
\begin{flushleft}
Katie R. Mollan,  Ilana M. Trumble, Sarah A. Reifeis, Orlando Ferrer, Camden P. Bay, Pedro L. Baldoni, and Michael G. Hudgens
\end{flushleft}
\begin{flushleft}
\textit\textit{Department of Biostatistics and Center for AIDS Research, The University of North Carolina, Chapel Hill, NC}
\end{flushleft}
\begin{flushleft}
Corresponding Author's Footnote: 
\end{flushleft}
\begin{flushleft}
Katie R. Mollan is a Senior Biostatistician at The University of North Carolina at Chapel Hill, 3126 McGavran-Greenberg Hall, CB \#7420, Chapel Hill, NC 27599 (email: kmollan@unc.edu)
\end{flushleft}

\section*{ABSTRACT}

Accurate power calculations are essential in small studies containing expensive experimental units or high-stakes exposures. Herein, exact power of the Wilcoxon Mann-Whitney rank-sum test \added{of a continuous variable} is formulated using a Monte Carlo approach and defining $P(X<Y)\equiv p$ \added{as a measure} of effect size, where $X$ and $Y$ denote random observations from two distributions hypothesized to be equal under the null. \added{Effect size $p$ fosters productive communications because researchers understand $p=0.5$ is analogous to a fair coin toss, and $p$ near 0 or 1 represents a large effect.} This approach is feasible even without background data. Simulations were conducted comparing the exact power approach to existing approaches by Rosner \& Glynn (2009), Shieh et al.\ (2006), Noether (1987), and O’Brien-Castelloe (2006). Approximations by Noether and O’Brien-Castelloe are \added{shown to be inaccurate} for small sample sizes. The Rosner \& Glynn and Shieh et al.\ approaches performed well in many small sample scenarios, though both are restricted to location-shift alternatives and neither approach is theoretically justified for small samples. The exact method is recommended and available in the R package \texttt{wmwpow}.

KEYWORDS: Mann-Whitney test, Monte Carlo simulation, non-parametric, power analysis, Wilcoxon rank-sum test 
 
\section*{1. Introduction}
Despite the current era of big data, there remains a practical need for power calculations of small preclinical, first-in-human, and basic science studies involving two independent samples. Accurate power calculations are critical when each experimental unit is expensive (e.g., macaques for preclinical HIV vaccine experiments) or the study is high stakes (e.g., novel HIV cure strategies where toxicity risks are unknown). Anti-conservative power approximations can result in an underpowered study and conservative approximations can lead to using more experimental units than necessary. Owing to small sample sizes, determining power in this setting is challenging because asymptotic approximations may not be reliable. An additional challenge common in many studies, such as preclinical or first-in-human trials, is the absence of relevant background data to inform power calculations.

In small studies with a continuous outcome (e.g., Kulkarni et al.\ 2011; Archin et al.\ 2014; Denton et al.\ 2014), the Wilcoxon Mann-Whitney (WMW) rank-sum test is often utilized to test for differences between groups (Wilcoxon 1945; Mann and Whitney 1947). Thus it is of interest to compute power of the WMW test against different alternatives.  Previous work on calculating power of the WMW test for a continuous outcome includes Haynam \& Govindarajulu (1966), Noether (1987), Collings \& Hamilton (1988), Lehmann (1998), Shieh et al.\ (2006), Zhao et al.\ (2008), Rosner \& Glynn (2009), and Divine et al.\ (2010). Power of the WMW test for ordered categorical outcomes was considered previously by Hilton \& Mehta (1993), Kolassa (1995), and Tang (2011, 2016).

In this paper, an exact approach for determining the power of the WMW test is formulated using Monte Carlo simulation. The approach is exact in that no asymptotic approximation is employed, and the amount of Monte Carlo error can be controlled by the user. In addition to being exact, an appealing aspect of this approach is that it can be implemented with or without background data. Effect size is defined by $p=P(X<Y)$, where $X$ and $Y$ denote random observations from the two distributions being compared. Equivalently, the effect size can  be expressed by the odds $p/(1-p)$ (O'Brien \& Castelloe 2006; Divine et al.\ 2013, 2017). Under a location-shift alternative, the WMW test null hypothesis is $p=0.5$, analogous to a fair coin toss. This effect size can be easily understood by collaborative investigators. Moreover, when background data are lacking, it can be more productive to discuss plausible values for $p$ with collaborators than to elicit parameterizations for each distribution or to quantify effect size using standard deviation units. Further, as shown here and by Rosner \& Glynn (2009), in many design scenarios the underlying distributions have minimal impact on power for a fixed effect size $p$.

The outline of the remainder of this paper is as follows. Section 2 presents several approaches to calculating power of the WMW test (with details in the Appendices). Section 3 presents simulation results comparing WMW test power calculations. Section 4 provides a motivating example, and Section 5 concludes with a discussion.

\section*{2. Methods}

Suppose $X_1,..., X_m$ and $Y_1,..., Y_n$ are independent identically distributed (iid) random variables with continuous cumulative distribution functions $F$ and $G$, respectively. It is of interest to test the null hypothesis $H_0: F=G$ versus the two-sided alternative hypothesis $H_A: F\neq G$. The WMW test statistic is $W= \sum_{i=1}^{m} \sum_{j=1}^{n} \varphi(Y_j - X_i)$ where $\varphi(Y_j - X_i)=1$ when $Y_j > X_i$, and 0 otherwise; i.e., the WMW statistic counts the number of times a $Y_j$ is larger than a $X_i$. Under $H_0$, the WMW statistic has mean $\mu_0=mn/2$ and variance $\sigma^2_0=mn(N+1)/12$ where $N=n+m$; as $m$ and $n$ tend to infinity, $(W-\mu_0)/\sigma_0$ has a limiting standard normal distribution under $H_0$ (Mann \& Whitney 1947).

Shieh et al.\ (2006) derived a large-sample approximation for power of the WMW test using the exact variance of $W$ under the alternative hypothesis $H_A$, and demonstrated that their approach was more accurate than the Noether (1987) and Lehmann (1998) approximations. Effect size in Shieh et al.\ was defined in terms of $G(x) = F(x-\theta)$, where $\theta$ is the location shift in the cumulative distribution function (CDF) and $H_0:\theta=0$. The Shieh et al.\ method is reformulated here using effect size $p$ (Appendix A) to facilitate interpretation and comparison to other approaches to estimating power of the WMW test. For large $m$ and $n$, power for the two-sided WMW test against a specific alternative hypothesis can be approximated by: 
\begin{equation} \label{eq:pow}
P\bigg\{\bigg|\frac{W-\mu_0}{\sigma_0} \bigg| >  z_{\alpha/2} \biggm| H_A\bigg\} \approx \Phi\Big(\frac{\mu - \mu_{0}-z_{\alpha/2}\sigma_{0}}{\sigma}\Big) + \Phi\Big(\frac{\mu_0 -\mu - z_{\alpha/2}\sigma_{0} }{\sigma}\Big) \
\end{equation}
where $\alpha$ is the significance level, $\Phi(\cdot)$ is the CDF of a standard normal distribution, $z_{\alpha/2} = \Phi^{-1}(1-\alpha/2)$, and $\mu$ and $\sigma$  are the mean and standard deviation of the WMW statistic under $H_A$, respectively. The mean under $H_A$, $\mu =mn/p$, depends upon effect size $p$, and the variance under $H_A$ can be expressed as: 
\begin{equation} \label{eq:ssig}
\sigma^2=mn\{p(1-p) + (n-1)(p_{2}-p^2) + (m-1)(p_{3}-p^2)\}
\end{equation}
where $\sigma^2$ depends upon effect size $p$ and underlying distributions $F$ and $G$ through $p_{2}$ and $p_{3}$ (Lehmann 1998; Shieh et al.\ 2006). 

Noether (1987) provides an approximation to the power of the WMW test which also relies on the normal approximation in Equation \ref{eq:pow}, but does not require selecting parametric models for $F$ and $G$. Instead, two additional assumptions are supposed: (i) $\sigma^2 = \sigma^2_0$, i.e., the variance of $W$ under $H_A$ is equal to the variance under $H_0$; and (ii) $N/(N+1) \approx 1$ (Appendix B). Assumptions (i) and (ii) may be dubious for small sample sizes. Clearly the approximation $N/(N+1) \approx 1$ only holds for large $N$. In addition, a study with small $m$ and $n$ will have adequate power only for large effect sizes, in which case $\sigma$ will not, in general, equal $\sigma_0$ (Shieh et al.\ 2006). 

Rosner and Glynn (2009) also provide a method for estimating the power of the WMW test which relies on the normal approximation in Equation \ref{eq:pow} but does not require selecting parametric models for $F$ and $G$.  Rosner and Glynn derive a closed-form estimate of power for location-shift alternatives defined after first applying a probit transformation to $F$ and $G$.

With modern computing, empirical (Monte Carlo) power calculation for the WMW test is feasible and accurate, particularly for small studies. As described below, empirical power computation entails repeated sampling from $F$ and $G$. Options for selecting $F$ and $G$ include: (i) specifying parametric distributions for both $F$ and $G$; (ii) specifying a parametric distribution for $F$ and choosing a value for $p$, which in turn imply a distribution for $G$; or (iii) resampling from a sufficient amount of background data (Collings \& Hamilton 1988; Hamilton \& Collings 1991). For studies where background data are unavailable or sparse, the resampling approach (iii) is not feasible. While approach (i) is feasible for small studies, it can be harder to interpret (e.g., presenting a mean difference in standard deviation units) compared to (ii) where one selects \added{effect size} $p$ or odds. Options (i) and (ii) are available in the R package described below.

The empirical method can provide power estimates that are effectively exact in practice. The general approach entails simulating multiple datasets from $F$ and $G$,  and computing the proportion of simulated datasets where the WMW test rejects the null. As the number of simulated datasets approaches $\infty$, empirical power converges in probability to the exact power of the WMW test. For a finite number of simulated data sets, the Monte Carlo error can be quantified, such that the number of simulations may be chosen to ensure this error is within an acceptable tolerance. In particular, let $Q$ be the number of rejections of $H_0$ among $S$ simulated datasets and let $p_q$ be the probability of rejecting $H_0$ with $Q \sim Binomial(S,p_q)$. For simulations under $H_0$, $p_q$ equals the type I error rate, and for simulations under a particular alternative hypothesis $H_A$, $p_q$ equals power. The power (or type I error) is estimated empirically by $\hat p_q = Q/S$. By the central limit theorem, for large $S$, $\hat p_q$ will be approximately normal with mean $p_q$ and the standard error of $\hat p_q$ will be no larger than $1/\sqrt[]{4S}$, which is $\approx 0.0016$ for $S=100,000$. This implies that $S=100,000$ simulated datasets will provide a precise power estimate to two decimal places. E.g., suppose $S=100,000$ and $Q=80,000$; then $\hat p_q=0.8$ and the corresponding 99\% Wald confidence interval (CI) for $p_q$ rounded to two decimal places is (0.80, 0.80). With $S=10,000$ the standard error of $\hat p_q$ is no larger than $\approx 0.005$, and for $Q/S=8,000/10,000$ the 99\% CI for $p_q$ is (0.79, 0.81).

The \texttt{wmwpow} R package provides three functions for estimating power: \texttt{wmwpowp}, \texttt{wmwpowd}, and \texttt{shiehpow}. For all three functions, the user inputs the sample sizes ($m,n$) and the significance level ($\alpha$). The function \texttt{wmwpowp} also takes inputs of the distribution for $F$ and the effect size $p$, and returns empirical power. For example, suppose the user inputs an exponential distribution with rate parameter $\mu$ for $F$ and a particular value for $p$; then \texttt{wmwpowp} solves for $G$. Available choices in \texttt{wmwpowp} for $F$ are the exponential, normal, and double exponential (Laplace) distributions, corresponding to the derivations in Appendix C. In each case, $F$ and $G$ are assumed to be in the same family or class of distributions; e.g., if $F$ is specified to be normal with mean $\mu_x$ and variance $\sigma_x^2$, then $G$ is assumed to be normal as well. If $F$ is exponential with rate $\mu$ and $p$ is fixed, then $G$ is completely specified. On the other hand, if $F$ is normal or double exponential and $p$ is fixed, then $G$ is not completely specified without additional assumptions. Therefore, for the normal and double exponential distributions, the function \texttt{wmwpowp} also takes as an input the scalar $k$ which specifies the ratio of standard deviations for $F$ and $G$. For $k=1$, choosing $p \neq 0.5$ corresponds to a location-shift alternative. Choosing $k \neq 1$ allows for unequal variances and thus a wider class of alternative hypotheses.

If specifying parametric distributions for both $F$ and $G$ is preferred, the function \texttt{wmwpowd} can be used to compute empirical power. \texttt{wmwpowd} allows the user to select from many standard continuous parametric distributions, including beta, exponential, normal, and Weibull. The function \texttt{wmwpowd} outputs the empirical power as well as the effect size $p$ and the equivalent odds corresponding to the $F$ and $G$ specified by the user.  

The \texttt{wmwpow} package also includes the function \texttt{shiehpow}, which implements the Shieh et al.\ method for location-shift alternatives assuming normal, shifted exponential, or double exponential distributions. The function \texttt{shiehpow} uses a shifted exponential distribution, whereas the exponential distribution in \texttt{wmwpowp} uses one rate parameter that defines both shape and location such that a common support [$0, \infty$) is maintained for $F$ and $G$.

\section*{3. Empirical Comparisons}

The performance of methods by Noether (1987), O'Brien-Castelloe (2006), Shieh et al.\ (2006), and Rosner \& Glynn (2009) were compared to empirical power results. Each method was formulated such that $\alpha$, $m$, $n$, and $p$ were the inputs, as well as an assumed probability distribution, when required. Power was estimated for effect size $p$ ranging from 0.50 to 0.95 by 0.05 (odds ranging from 1 to 19).

The approach of Shieh et al.\ was implemented using the R package \texttt{wmwpow}, function \texttt{shiehpow} with the formulae shown in Appendix A. The Noether approach (Appendix B) was also implemented in R. The O'Brien-Castelloe approach was applied using the SAS Power procedure (\textit{twosamplewilcoxon}, SAS/STAT v14.2); default settings were used and distributional assumptions were $X \sim N(0,1)$ and $Y\sim N(\mu_y,1)$, solving for $\mu_y$ by inputting values of $p$ into the equation shown in Appendix C.2. Rosner \& Glynn (2009) provided a SAS macro (\textit{\%WilcxPowerContinuousNties}) for their approach. Empirical power was computed as the proportion of rejections of $H_0$ under a specific alternative hypothesis over $S$ simulated datasets; $S=100,000$ simulated datasets were used for $n,m<20$, and $S=10,000$ simulated datasets for $n,m\geq20$. Computations were conducted in R version 3.4.3 and SAS version 9.4 (Cary, NC).

\begin{figure}
  \centering
   \subfloat[\added{Exact power by sample size}]{\includegraphics[width=0.5\textwidth]{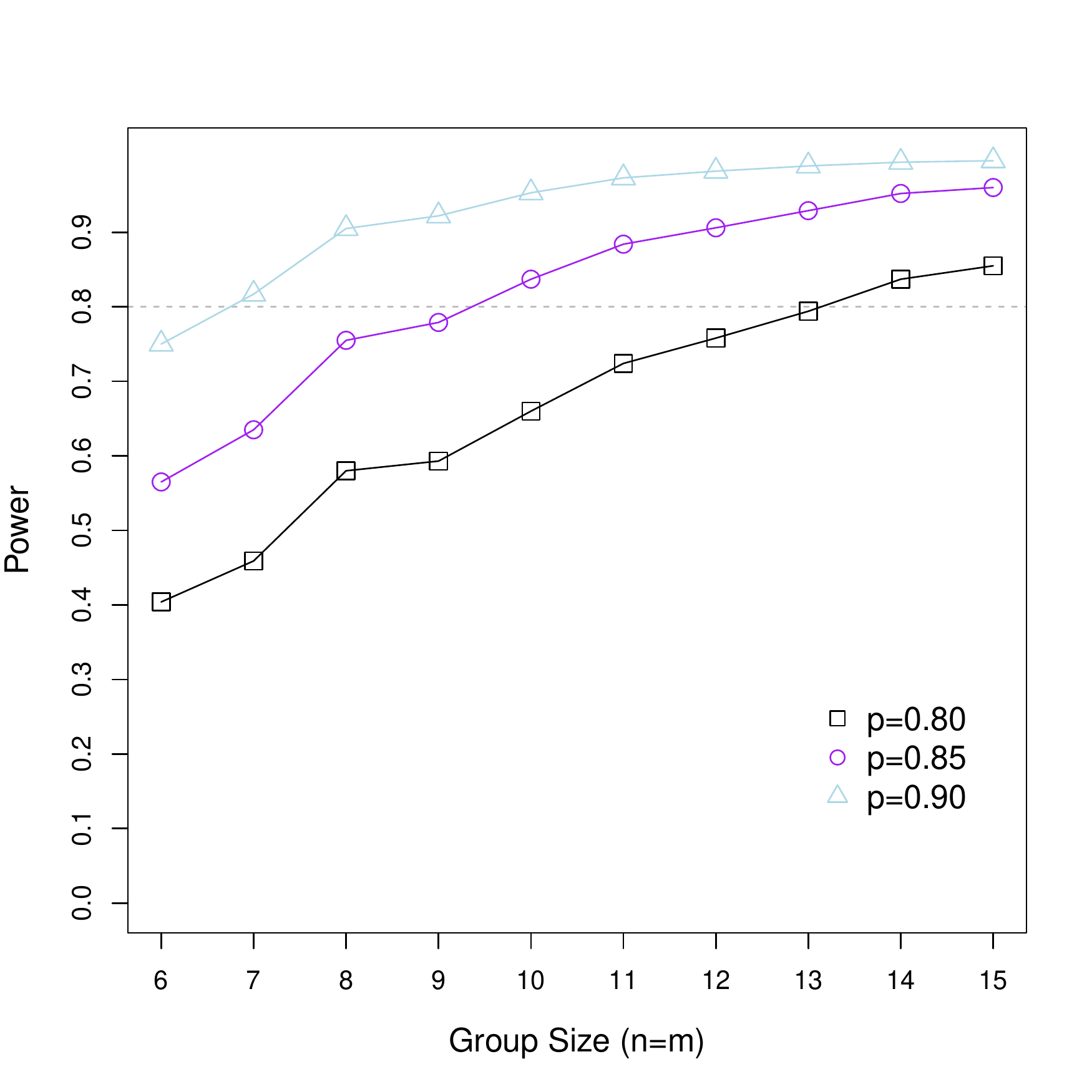}\label{fig:fpowvsn}} 
  \hfill
  \subfloat[$n=m=6$]{\includegraphics[width=0.5\textwidth]{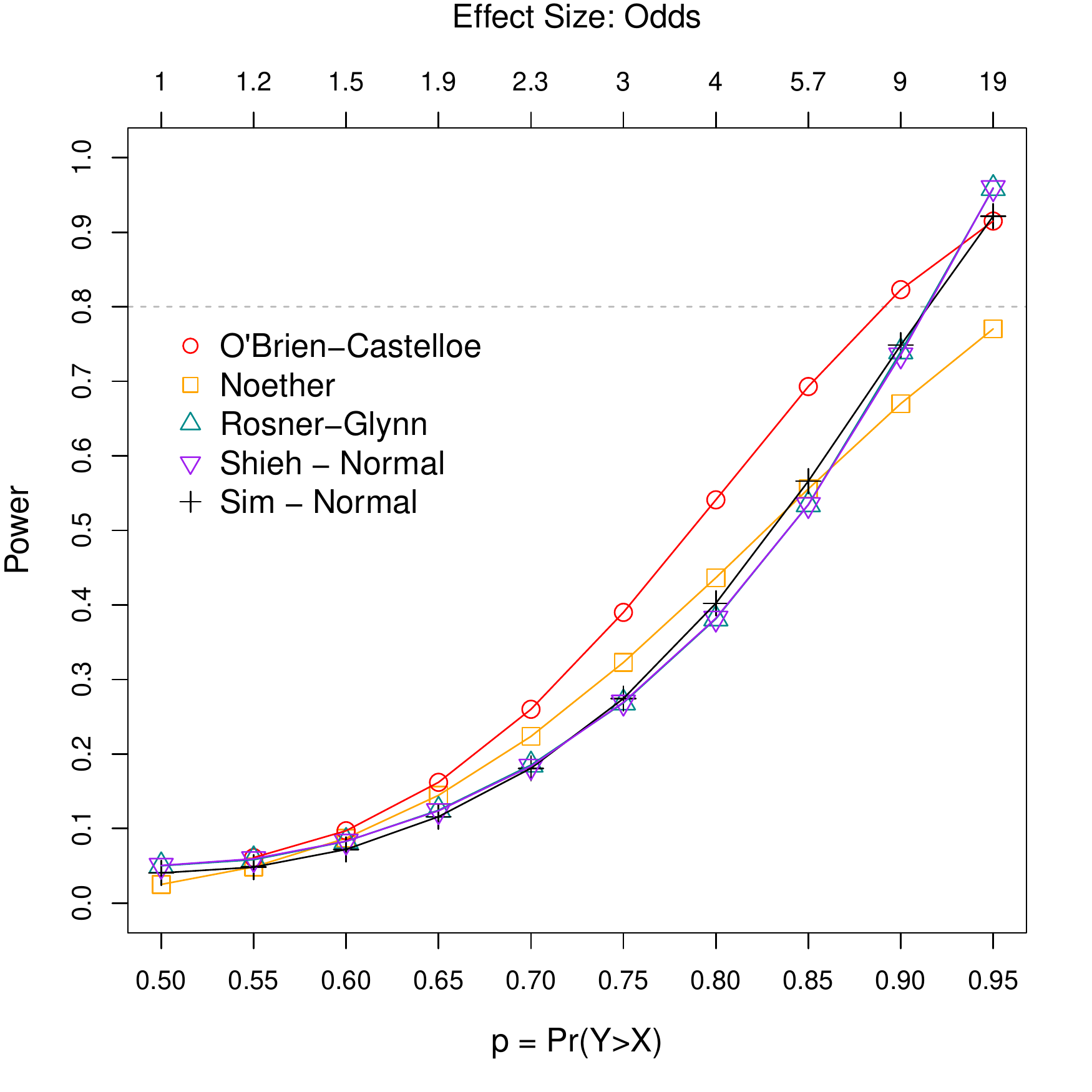}\label{fig:f1a}}
  \hfill
 \subfloat[$n=m=15$]{\includegraphics[width=0.5\textwidth]{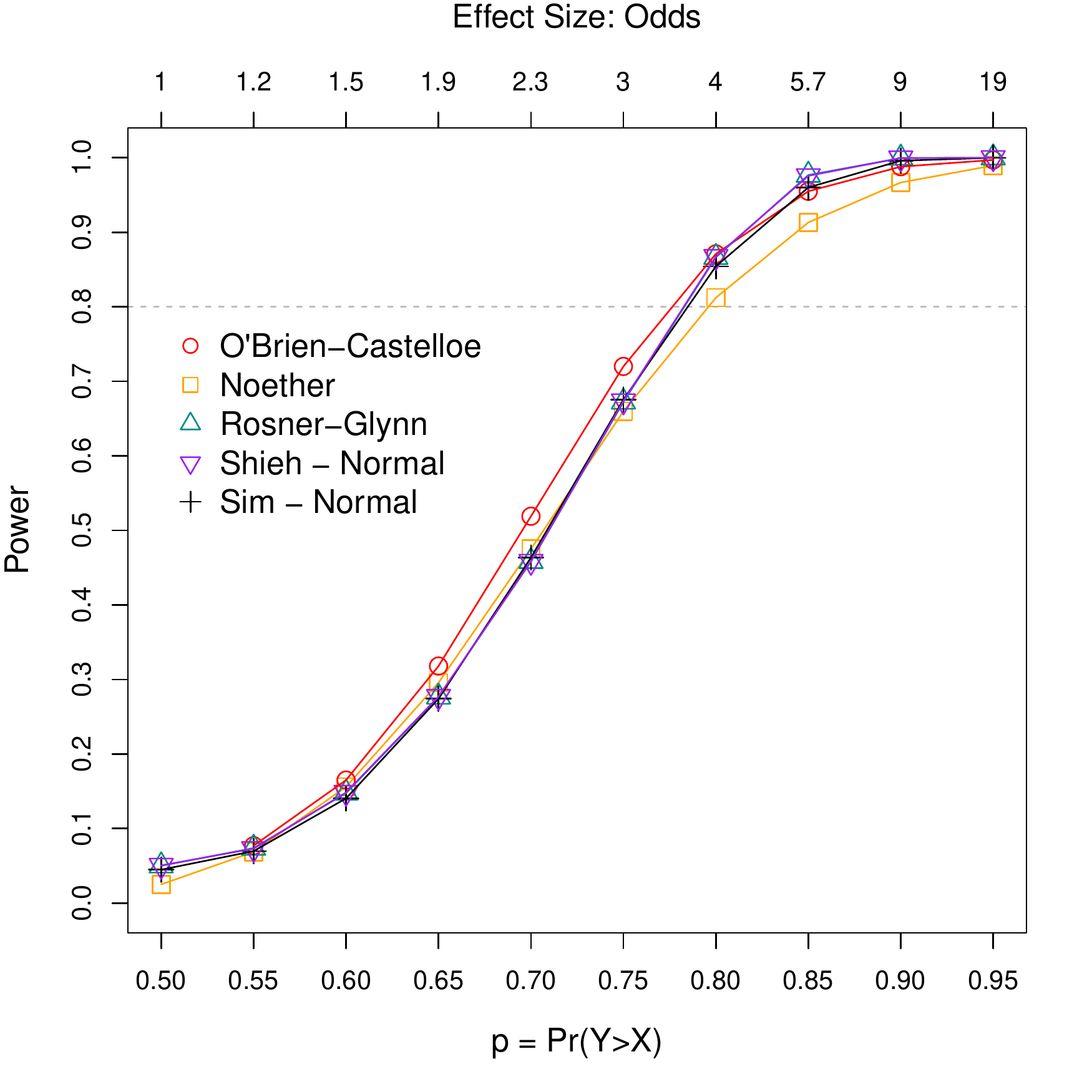}\label{fig:f1b}}
   \hfill
   \subfloat[$n=m=50$]{\includegraphics[width=0.5\textwidth]{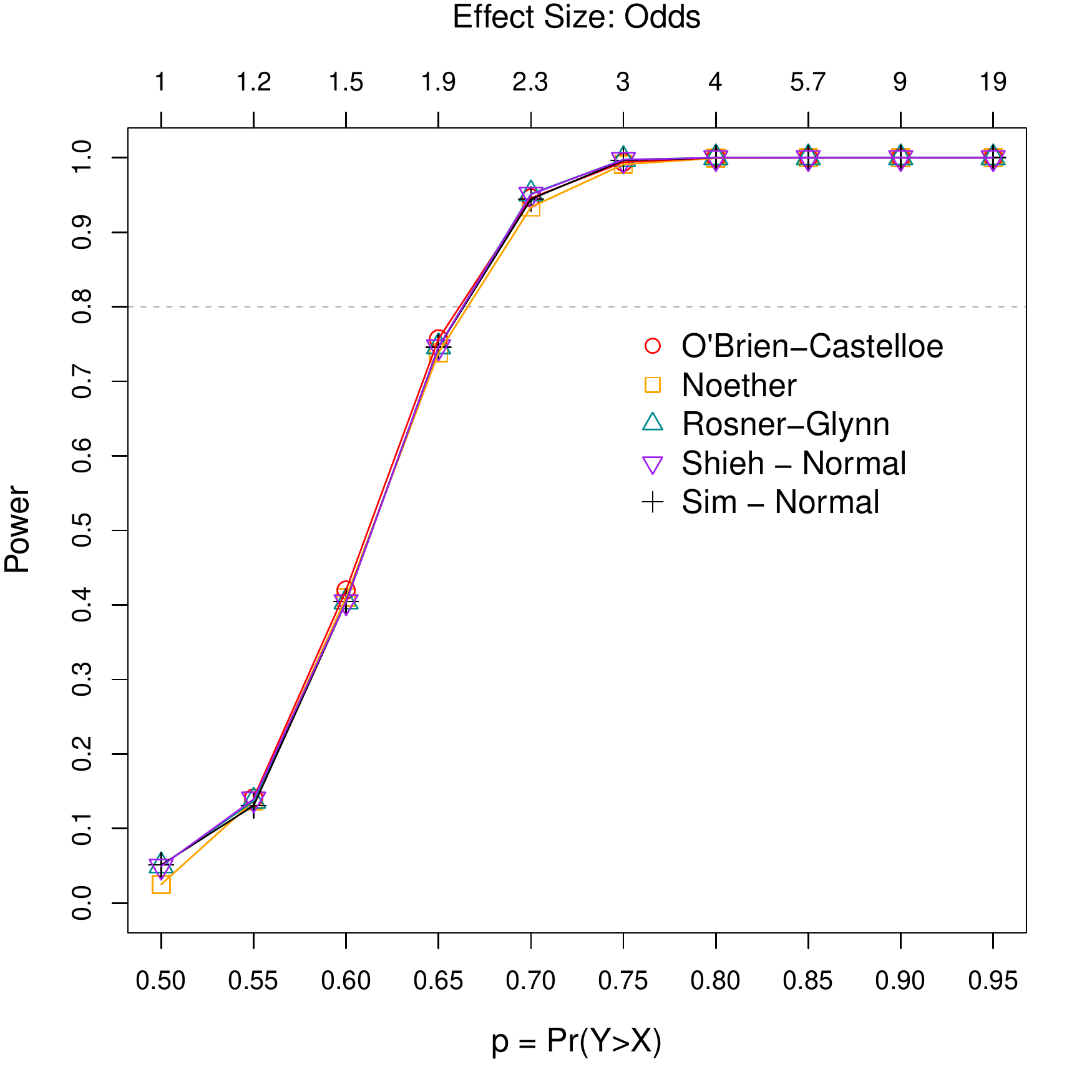}\label{fig:f50}}
   \hfill
   
  \caption{Power and type I error for the 2-sided Wilcoxon Mann-Whitney rank-sum test, $\alpha=0.05$. Sim = empirical simulation. \added{Panel (a) presents exact power from the empirical method with a normal distribution and equal standard deviations ($k=1$), and panels (b-d)} compare power results for a given sample size per group.}
\end{figure}

\added{Exact power for large effect sizes ($p \geq 0.8$) over a range of small samples sizes ($n=m=6$ to 15) is shown in Figure \ref{fig:fpowvsn} as calculated using \texttt{wmwpowp}.} Comparisons between the empirical power calculations and results from Shieh et al., Rosner-Glynn, Noether, and the O'Brien-Castelloe methods are shown in Table \ref{tab:tab1} and Figure \ref{fig:f1a}-\ref{fig:f50}. For $m=n=6$ per group, the Shieh et al.\ and Rosner-Glynn methods provided very similar results (Figure \ref{fig:f1a}). For a given $p$, varying the distributions for $F$ and $G$ had negligible effect on the power. The O'Brien-Castelloe approximation was typically anti-conservative for small $m$ and $n$ (e.g., $m=n=6$). The Noether approximation was both anti-conservative or over-conservative depending upon effect size $p$ and sample sizes (Figures \ref{fig:f1a} and \ref{fig:f1b}). As $m$ and $n$ increase, power results from the methods evaluated here became increasingly similar, as expected. For $m,n \geq 50$, all of the methods yielded similar results (Figure \ref{fig:f50}). 

Generally, the Shieh et al.\ and Rosner-Glynn approaches tended to well approximate exact (empirical) power. However, for small unequal sample sizes (e.g., $m=6, n=12$), the Shieh et al.\ and Rosner-Glynn power estimates can differ, as demonstrated in the bottom of Table \ref{tab:tab1}. Note the Rosner-Glynn approach gives the same power estimate when $m=6,n=12$ and $m=12,n=6$ for a fixed effect size $p$. In contrast, Shieh et al.\ power estimates need not be the same when the values of $m$ and $n$ are interchanged as can be seen from Equation \ref{eq:ssig} and Appendix A ($p_{2}$ and $p_{3}$ are unequal for non-symmetric distributions).

\begin{figure}[!tbp]
\centering

\subfloat[$n=m=6$]{\includegraphics[width=0.5\textwidth]{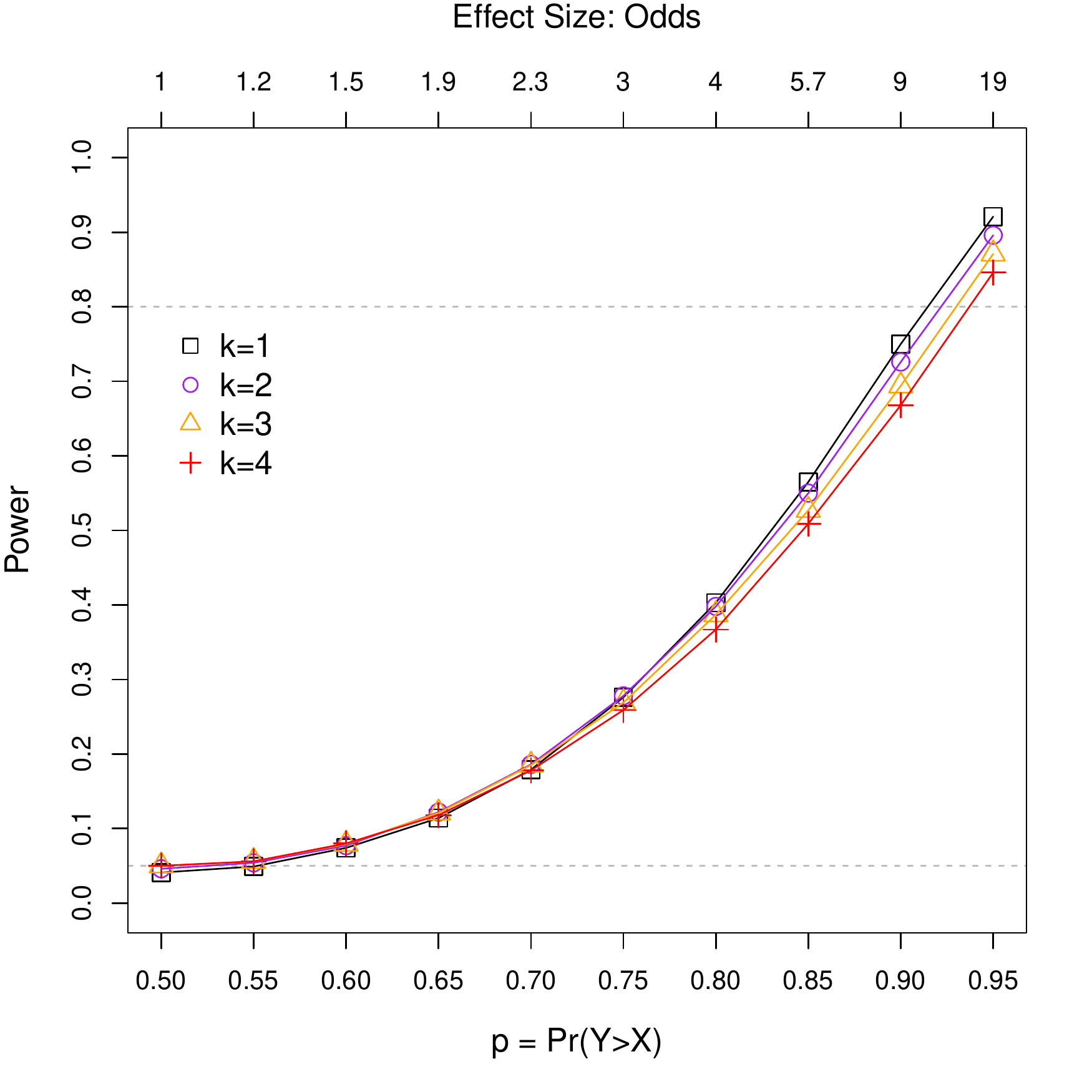}\label{fig:fknorm}}
\hfill
\subfloat[$n=m=15$]{\includegraphics[width=0.5\textwidth]{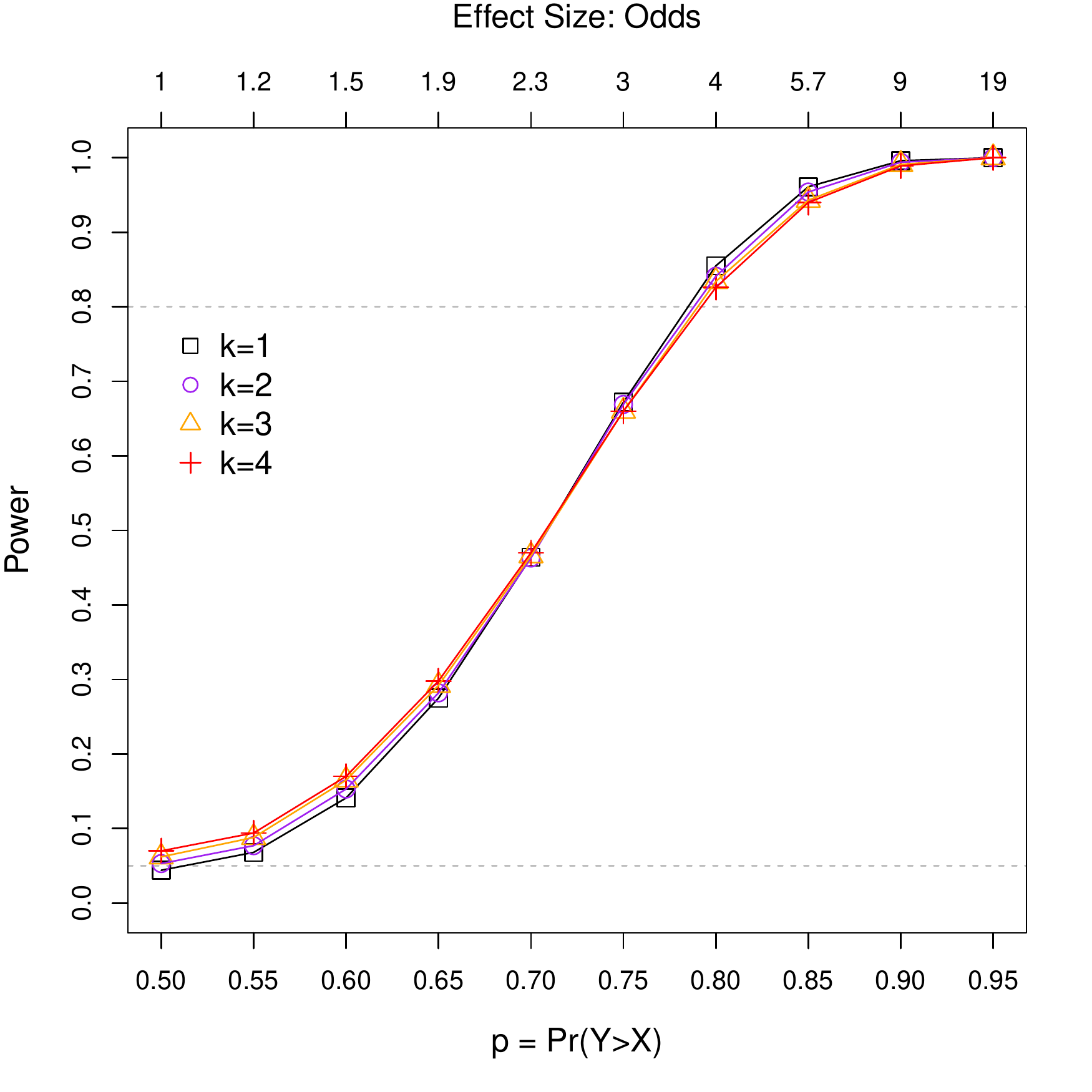}\label{fig:fknorm15}}

\caption{Empirical power for $F$ and $G$ normal with standard deviation ratio $k= \sigma_y / \sigma_x=1, 2, 3, 4$, $\alpha=0.05$.}
\end{figure}

Empirical power for alternative hypotheses where $F$ and $G$ are normal with unequal variances is shown in Figure \ref{fig:fknorm}. For $m=n=6$, power decreases as the degree of variance heterogeneity increases (i.e., as $k$ increases).  Varying $k$ had less impact for $m=n=15$. Note that if $k \neq 1$, then the null hypothesis $H_0: F = G$ does not hold even if $p=0.5$. Hence, in Figure \ref{fig:fknorm15} the empirical power is above $\alpha=0.05$ for $p=0.5$ and $k=3,4$.

\begin{table}[ht]
\centering
\caption{\label{tab:tab1}Power and type I error for the 2-sided Wilcoxon Mann-Whitney rank-sum test, $\alpha=0.05$}
\label{my-label}
\begin{tabular}{l||r|rrrrrr}
\hline
\multicolumn{1}{c||}{Sample Size} & \multicolumn{1}{c|}{Method} & \multicolumn{6}{c}{Effect Size}                           \\
                                  &                             & p=0.5 & p=0.7 & p=0.75 & p=0.8 & p=0.85 & p=0.9            \\ \hline
n=6, m=6                          & Noether                     & 3     & 22    & 32     & 44    & 56     & 67               \\
                                  & O'Brien-Castelloe           & n/a   & 26    & 39     & 54    & 69     & 82               \\
                                  & Rosner-Glynn                & 5     & 19    & 27     & 38    & 53     & 74               \\
                                  & Empirical - Normal          & 4     & 18    & 28     & 40    & 56     & 75               \\
                                  & Empirical - Exp             & 4     & 18    & 28     & 40    & 56     & 74               \\
                                  & Empirical - Laplace         & 4     & 18    & 28     & 39    & 55     & 72               \\
                                  & Shieh - Normal              & 5     & 18    & 27     & 38    & 53     & 74               \\
                                  & Shieh - Shifted Exp         & 5     & 19    & 28     & 39    & 53     & 72               \\
                                  & Shieh - Laplace             & 5     & 19    & 27     & 38    & 53     & 72               \\ \hline
n=15, m=15                        & Noether                     & 3     & 48    & 66     & 81    & 91     & 97               \\
                                  & O'Brien-Castelloe           & n/a   & 52    & 72     & 87    & 96     & 99               \\
                                  & Rosner-Glynn                & 5     & 46    & 67     & 87    & 98     & \textgreater{}99 \\
                                  & Empirical - Normal          & 5     & 47    & 67     & 85    & 96     & \textgreater{}99 \\
                                  & Empirical - Exp             & 5     & 46    & 68     & 86    & 96     & \textgreater{}99 \\
                                  & Empirical - Laplace         & 5     & 46    & 68     & 85    & 95     & 99               \\
                                  & Shieh - Normal              & 5     & 46    & 67     & 86    & 98     & \textgreater{}99 \\
                                  & Shieh - Shifted Exp         & 5     & 46    & 67     & 85    & 97     & \textgreater{}99 \\
                                  & Shieh - Laplace             & 5     & 46    & 67     & 86    & 97     & \textgreater{}99 \\ \hline
n=6, m=12                         & Rosner-Glynn                & 5     & 25    & 37     & 53    & 73     & 92               \\
                                  & Empirical - Exp             & 4     & 24    & 37     & 54    & 73     & 90               \\
                                  & Shieh - Shifted Exp         & 5     & 23    & 36     & 54    & 74     & 93               \\ \hline
n=12, m=6                         & Rosner-Glynn                & 5     & 25    & 37     & 53    & 73     & 92               \\
                                  & Empirical - Exp             & 4     & 26    & 39     & 55    & 72     & 86               \\
                                  & Shieh - Shifted Exp         & 5     & 27    & 39     & 53    & 69     & 86               \\ \hline
\end{tabular}
\caption*{\added{Above, $n$ and $m$ are the two sample sizes. $p$ is a measure of effect size, i.e., the probability that the first random variable is less than the second variable, $P(Y>X)$. Statistical power (\%) is displayed rounded to the nearest whole percentage. Effect size $H_0: p=0.5$ presents type I error. Exp=Exponential}}
\end{table}

\section*{4. Motivating Example}

Consider a proposed study of $m=n=15$ per group with the sample size limited by ethical (e.g., safety), recruitment, or budgetary constraints. Given the limited feasible sample size, an accurate assessment of power is crucial for deciding whether the study should proceed as planned. In this study, background data on the outcome are dearth, and yet power calculations are still needed if null hypothesis significance testing is planned. In some cases, the study team should change the study design to focus on estimation and collection of pilot data without hypothesis testing. Here we proceed assuming that group comparisons are essential to the study objectives.

For example, in early phase clinical trials evaluating potential cures for HIV, sample sizes are typically limited to mitigate potential risks to participants. An outcome of interest, HIV replication index, is a relatively new measure used in HIV cure research with limited background data. \added{Replication index is defined as the mean number of live daughter cells created by each parent cell over a specific length of time; this provides proliferative capacity on a per-cell basis, independent of the number of cells that originally started to proliferate (Clutton et al.\ 2016), and is closely related to proliferation index (Migueles et al.\ 2009).} 

Suppose the investigators choose a 0.05 significance level and decide $p=0.8$ or larger is a meaningful effect size, i.e., an 80\% or larger true probability that the HIV replication index for any given individual in the placebo group is higher than for any given individual in the treatment group. Assuming $p=0.8$ (or equivalently, a true odds of 4 or larger) and using an empirical power approach, thirty individuals ($m=n=15$ per group) will provide 85\% power to detect a difference between two independent groups (placebo versus treatment). Rosner-Glynn, and  O'Brien-Castelloe power estimates were both 87\% and  the Shieh et al. estimate was 86\%, whereas the Noether approximation was conservative in this example (81\% power). Empirical power for the exact 2-sided WMW test was conducted assuming a normal distribution for $\log_{10}$ replication index and 100,000 datasets of size $m=n=15$ were generated. 

\section*{5. Discussion}

Empirical power calculation is accurate and feasible for many power scenarios including small sample settings, unequal variance, and unequal group sample sizes. The power approximations of Noether and O’Brien-Castelloe are not reliably accurate for small sample sizes. The Rosner \& Glynn and Shieh et al.\ approaches performed well in many small sample scenarios, though both are restricted to location-shift alternatives and neither approach is theoretically justified to provide accurate power estimates for small samples. In contrast, the empirical power approach can evaluate a wider class of alternative hypotheses and is valid for any sample size.

In some settings, it may be anticipated that ties will occur in the observed data. Ties can arise when the underlying variable is continuous, but the variable is measured or recorded with limited granularity such that two or more individuals may have the same recorded value. Estimating power of the WMW test when ties may or may not be present was not considered here; it is often not practical to ascertain the \textit{a priori} probability of a tie occurring. Zhao et al.\ (2008) generalized the Noether (1987) method to handle ties, making the assumption that the variance of the test statistic $W$ under the alternative $H_A$ is the same as under $H_0$; this assumption may be dubious when group sample sizes are small. If adequate background information is available regarding ties, one can simulate data accordingly (e.g., resample from the background data), and proceed with empirical power calculation. Ordered categorical data can be thought of as an extreme case of ties, and can be simulated directly using category probabilities (e.g., the tabled distribution within the SAS function \texttt{RAND}). WMW test power calculation for ordered categorical data is also available in StatXact software (Hilton \& Mehta 1993).

Exact power calculation via Monte Carlo simulation is recommended whenever computationally feasible. Empirical power calculation for the rank-sum test is available in the commercial software PASS by inputting parametric distributions for $F$ and $G$. However, PASS version 16 does not yet provide $p$ or odds as an input or output value. The R package \texttt{wmwpow} can be used to compute empirical power with either $p$ or odds as an input (or alternatively $F$ and $G$), and is free and publicly available on CRAN. 

\section*{Acknowledgements}
We thank Genevieve Clutton, Kristina De Paris, and J.\ Victor Garcia-Martinez and the UNC HIV research community for requesting power calculations that motivated this work. We also thank the Editor, Associate Editor, two reviewers, and Nader Gemayel for their helpful comments and suggestions, \added{and Marion McPhee and Bernard Rosner for providing an updated SAS macro}.

\section*{Funding}
This research was supported by the University of North Carolina at Chapel Hill Center for AIDS Research (CFAR), an NIH funded program P30 AI50410. 

\section*{Appendices}
\section*{Appendix A}
As shown in Lehmann (1998) and Shieh et al.\ (2006), the variance $\sigma^2$ of the WMW statistic under $H_A$ depends on $F$ and $G$; this dependence can be formulated using $p_{2}$ and $p_{3}$ for a location-shift alternative, with $p_{2}=p_{3}$ when distributions $F$ and $G$ are symmetric. When the underlying distributions of $F$ and $G$ are \added{shifted} exponential \added{(a non-symmetric distribution)}, $\theta = -\ln[2(1-p)]$ for $p$ in (0.5,1), $p_{2}=1-2/3e^{-\theta}$, and $p_{3}= 1-e^{-\theta} + 1/3e^{-2\theta}$. When the distributions of $F$ and $G$ are double exponential (Laplace), $\theta = -L(4(p - 1)/e^{2})$ where $L$ is the Lambert-W function used to solve for $x$ when $y=xe^{x}$ and $p_{2}=p_{3} = 1-(7/12 + \theta/2)e^{-\theta} -1/12e^{-2\theta}$. Lastly, for the normal case, $F \sim N(0,1)$, $\theta = \sqrt[]{2}\Phi^{-1}(p)$ and $p_{2}=p_{3}= E[\{\Phi(Z+\theta)\}^2], \textrm{where} \space \ Z \sim N(0,1)$. 
\section*{Appendix B}

Noether (1987) provided an approximation to the power of the WMW test assuming $\sigma = \sigma_0$, and $N/(N+1) \approx 1$, where $N=m+n$. Consider a one-sided WMW test, in which case the power equals:
$$1-\beta = P\bigg(Z > \frac{\mu_0 - \mu}{\sigma} + \frac{z_\alpha \sigma_0}{\sigma}\bigg).$$

Let $c=m/N$ and $z_{\beta} = \Phi^{-1}(1-\beta)$. Then under the assumption $\sigma = \sigma_0$, it follows that
$$\Big(\frac{\mu_0 - \mu}{\sigma_0}\Big)^2=\frac{12N^2 c(1-c)(p- 0.5)^2}{N+1} = (z_\alpha + z_\beta)^2$$
or equivalently
$$\frac{N^2}{N+1} = \frac{(z_\alpha + z_\beta)^2}{12c(1-c)(p-0.5)^2}.$$ 
Assuming $N/(N+1) \approx 1$, it follows that
\[N\approx \frac{(z_\alpha + z_\beta)^2}{12c(1-c)(p-0.5)^2}\]
and therefore power of the WMW test is approximated by
\[1-\beta \approx \Phi\Big[\sqrt{12Nc(1-c)(p-0.5)^2} - z_{\alpha} \Big].\]

\section*{Appendix C}

Consider the general form for $p = P(X<Y) = \int_{-\infty}^{\infty} \int_{-\infty}^{y} f_X(x) g_Y(y)dxdy=\int_{-\infty}^{\infty} g_Y(y) F_X(y)dy$, where $f_X(x)$ and $g_Y(y)$ are probability density functions for $X$ and $Y$, respectively. The following three distributions are implemented in the R package \texttt{wmwpow}, function \texttt{wmwpowp}.

\subsection*{C.1: Exponential}
Let $X\sim Exp(\mu)$ and $Y\sim Exp(\lambda)$, where $\mu$ and $\lambda$ are exponential rate parameters. Then $p = P(X<Y) = \int_{0}^{\infty} \int_{x}^{\infty} \mu \lambda e^{-\mu x} e^{-\lambda y} dydx = \mu/(\lambda+\mu)$, and therefore $\lambda = \mu(1-p)/p$. 

\subsection*{C.2: Normal}
Let $X\sim N(\mu_x, \sigma^2_x)$ and $Y\sim N(\mu_y, \sigma^2_y)$ such that $X-Y \sim N(\mu_x - \mu_y, \sigma^2_x+\sigma^2_y)$. This implies $p = P(X-Y<0) = \Phi\Big(\frac{\mu_y - \mu_x}{\sqrt[]{\sigma^2_x+\sigma^2_y}}\Big)$, and therefore $\mu_y = \mu_x + \Phi^{-1}(p)\sqrt[]{\sigma^2_x+\sigma^2_y}$.

\subsection*{C.3: Double Exponential}

Let $X\sim Laplace(\mu_x, \sigma_x)$ and $Y\sim Laplace(\mu_y, \sigma_y)$. Then $\mu_y$ can be found as follows. Recall the cumulative distribution function of a Laplace random variable is\[
  F_X(x) =
  \begin{cases}
                                   \frac{1}{2} e^{\frac{x-\mu_x}{\sigma_x}} & \text{if $x \leq \mu_x$} \\
                                   1-\frac{1}{2} e^{-\frac{x-\mu_x}{\sigma_x}} & \text{if $x > \mu_x$.} \\
  \end{cases}
\]
This implies $p=P(X<Y)$ \[ 
= \int_{-\infty}^{\mu_x} \bigg(\frac{1}{2} e^{\frac{y-\mu_x}{\sigma_x}}\bigg) \bigg(\frac{1}{2\sigma_y} e^{-\frac{|y-\mu_y|}{\sigma_y}}\bigg)dy + \int_{\mu_x}^{\infty} \bigg(1-\frac{1}{2} e^{-\frac{y-\mu_x}{\sigma_x}}\bigg) \bigg(\frac{1}{2\sigma_y} e^{-\frac{|y-\mu_y|}{\sigma_y}}\bigg)dy.\]

Thus $\mu_y$ can be found by solving numerically \[
\Bigg[\int_{-\infty}^{\mu_x} \bigg(\frac{1}{2} e^{\frac{y-\mu_x}{\sigma_x}}\bigg) \bigg(\frac{1}{2\sigma_y} e^{-\frac{|y-\mu_y|}{\sigma_y}}\bigg)dy + \int_{\mu_x}^{\infty} \bigg(1-\frac{1}{2} e^{-\frac{y-\mu_x}{\sigma_x}}\bigg) \bigg(\frac{1}{2\sigma_y} e^{-\frac{|y-\mu_y|}{\sigma_y}}\bigg)dy\Bigg] - p = 0\] 

using any standard one-dimensional root finding method given $p$, $\mu_x$, $\sigma_x$, and $\sigma_y$.

\section*{References}

Archin, N. M., Bateson, R., Tripathy, M. K., et al.\ (2014). HIV-1 expression within resting CD4+ T cells after multiple doses of vorinostat. The Journal of Infectious Diseases, 210(5), 728-735.

\added{Clutton G, Xu Y, Baldoni PL, et al.\ (2016). The differential short-and long-term effects of HIV-1 latency-reversing agents on T cell function. Scientific Reports, 6:30749.}

Collings, B. J., and Hamilton, M. A. (1988). Estimating the power of the two-sample Wilcoxon test for location shift. Biometrics, 44(3), 847-860.

Denton, P. W., Long, J. M., Wietgrefe, S. W., et al.\ (2014). Targeted cytotoxic therapy kills persisting HIV infected cells during ART. PLoS Pathogens, 10(1), e1003872.

Divine, G., Kapke, A., Havstad, S., and Joseph, C. L. (2010). Exemplary data set sample size calculation for Wilcoxon–Mann–Whitney tests. Statistics in Medicine, 29(1), 108-115.

Divine, G., Norton, H. J., Hunt, R., and Dienemann, J. (2013). A review of analysis and sample size calculation considerations for Wilcoxon tests. Anesthesia \& Analgesia, 117(3), 699-710.

Divine, G. W., Norton, H. J., Barón, A. E., and Juarez-Colunga, E. (2018). The Wilcoxon-Mann-Whitney procedure fails as a test of medians. The American Statistician, 72(3), 278-286.

Hamilton, M. A., and Collings, B. J. (1991). Determining the appropriate sample size for nonparametric tests for location shift. Technometrics, 33(3), 327-337.

Haynam, G. E., and Govindarajulu, Z. (1966). Exact power of Mann-Whitney test for exponential and rectangular alternatives. The Annals of Mathematical Statistics, 37(4), 945-953.

Hilton, J. F., and Mehta, C. R. (1993). Power and sample size calculations for exact conditional tests with ordered categorical data. Biometrics, 49(2), 609-616.

Kolassa, J. E. (1995). A comparison of size and power calculations for the Wilcoxon statistic for ordered categorical data. Statistics in Medicine, 14(14), 1577-1581.

Kulkarni, S., Savan, R., Qi, Y., et al.\ (2011). Differential microRNA regulation of HLA-C expression and its association with HIV control. Nature, 472(7344), 495-498.

Lehmann, E.L. (1998). Nonparametrics: Statistical Methods Based on Ranks, Upper Saddle River, New Jersey: Prentice Hall.

Mann, H. B., and Whitney, D. R. (1947). On a test of whether one of two random variables is stochastically larger than the other. The Annals of Mathematical Statistics, 18(1), 50-60.

\added{Migueles SA, Weeks KA, Nou E, et al.\ (2009). Defective human immunodeficiency virus-specific CD8+ T-cell polyfunctionality, proliferation, and cytotoxicity are not restored by antiretroviral therapy. Journal of Virology, 83(22), 11876-89.}

Noether, G. E. (1987). Sample size determination for some common nonparametric tests. Journal of the American Statistical Association, 82(398), 645-647.

O’Brien R. G., Castelloe J. M. (2006). Exploiting the link between the Wilcoxon-Mann-Whitney test and a simple odds statistic. In Proceedings of the Thirty-first Annual SAS Users Group International Conference, Paper 209–31. SAS Institute Inc: Cary, NC.

Rosner, B., and Glynn, R. J. (2009). Power and sample size estimation for the Wilcoxon rank sum test with application to comparisons of C statistics from alternative prediction models. Biometrics, 65(1), 188-197.

Shieh, G., Jan, S. L., Randles, R. H. (2006). On power and sample size determinations for the Wilcoxon–Mann–Whitney test. Journal of Nonparametric Statistics, 18(1), 33-43.

Tang, Y. (2011). Size and power estimation for the Wilcoxon–Mann–Whitney test for ordered categorical data. Statistics in Medicine, 30(29), 3461-3470.

Tang, Y. (2016). Notes on Kolassa’s method for estimating the power of Wilcoxon–Mann–Whitney test. Communications in Statistics-Simulation and Computation, 45(1), 240-251.

Wilcoxon, F. (1945). Individual comparisons by ranking methods. Biometrics Bulletin, 1(6), 80-83.

Zhao, Y. D., Rahardja, D., and Qu, Y. (2008). Sample size calculation for the Wilcoxon–Mann–Whitney test adjusting for ties. Statistics in Medicine, 27(3), 462-468.

\end{document}